\documentclass[sigconf]{acmart}

\AtBeginDocument{%
  }

\copyrightyear{2026}
\acmYear{2026}
\setcopyright{cc}
\setcctype{by}
\acmConference[MUM '26]{25th International Conference on Mobile and Ubiquitous Multimedia}{December 01--04, 2026}{Glasgow, United Kingdom}
\acmBooktitle{25th International Conference on Mobile and Ubiquitous Multimedia (MUM '26), December 01--04, 2026, Glasgow, United Kingdom}
\acmDOI{10.1145/3861188.3861215}
\acmISBN{979-8-4007-2791-7/2026/12}

\usepackage{subcaption}
\usepackage{enumitem}

\begin{document}
\title[Exploring a Wearable-Mediated Channel
for Discreet Human-Agent Interaction]
{Just Between Us? Exploring a Wearable-Mediated Channel
for Discreet Human-Agent Interaction in a Shared Game Setting}


\author{Atikkhan Faridkhan Nilgar}
\affiliation{%
  \institution{Universit{\"a}t Siegen}
 \city{Siegen}
   \state{NRW}
   \country{Germany}}
\email{atikkhan.nilgar@uni-siegen.de}
\orcid{0009-0001-4948-614X}

\author{Manuel Dietrich}
\affiliation{%
  \institution{Honda Research Institute Europe GmbH}
 \city{Offenbach am Main}
   \country{Germany}
}
\email{manuel.dietrich@honda-ri.de}
\orcid{0000-0001-6819-8656}

\author{Kristof Van Laerhoven}
\affiliation{%
  \institution{Universit{\"a}t Siegen}
\city{Siegen}
   \state{NRW}
   \country{Germany}
}
\email{kvl@eti.uni-siegen.de}
\orcid{0000-0001-5296-5347}

\renewcommand{\shortauthors}{Nilgar et al.}


\begin{abstract}
In shared settings, social robots may need to exchange information with an individual while limiting how readily co-present others can access or interpret that information.
Publicly observable speech and gestures do not necessarily
support this need. We therefore investigate a wearable-mediated channel for
discreet interaction, combining in-ear audio messages with
deliberate rightward head tilts as responses. To examine its feasibility
and participants' experiences, we studied 20 pairs
(40 participants) playing a card game with a screen-based
conversational agent representing a social robot. A researcher
controlled this robot character in a Wizard-of-Oz setup.
Each pair played with and without the channel: one participant
wore the device (wearer), while the other (non-wearer) learned
about it after gameplay. The channel provided information
that could help the wearer or robot character choose an
opponent they could defeat. This exchange and the resulting opponent selection occurred as intended in 39 of 40 channel-mediated turns. Wearers reported low perceived
pressure/tension in both conditions, with no statistically
significant difference, and favorable immediate usability after
use. However, retrospective interviews revealed differing
accounts of noticing among non-wearers and contrasting
perceptions of teamwork and unfairness. These findings support the exchange's feasibility
within this game and highlight awareness and fairness as
considerations for future designs.
\end{abstract}





\maketitle

\section{Introduction}

Social interactions involving multiple people are common in everyday life. As social robots become increasingly integrated into society, they need to interact effectively in multi-party or group settings \cite{Nigro2025SocialGroup,Sebo2020RobotsInGroups}. Robots typically communicate in these settings through speech, gestures, or body movements that everyone present can observe. Although this overt communication can support transparency and mutual understanding, it may be unsuitable when information is intended for only an individual, particularly when that information is privacy-sensitive \cite{Nilgar2024User,Nilgar2025Privacy,Dietrich2023Whatshould,Tang2022Confidant,Zand2025SensitiveTelepresence}. For example, a healthcare robot may need to coordinate with a patient, healthcare provider, or caregiver without unnecessarily exposing patient information to others who are present \cite{jayaraman2024social}. This creates a need for an interaction channel that can direct task-relevant information to a person and receive a response while limiting how readily co-present others can access or interpret the exchange.

Humans address related communication needs through combinations of explicit and implicit verbal and gestural cues \cite{Shah2010AnEmpirical}. Robots can likewise use nonverbal cues such as gaze or body orientation to direct attention \cite{Admoni2017Socialeyegaze,Mutlu2009FootingIH,Mutlu2009NonverbalLI}. Although these cues can indicate an intended recipient, they remain publicly observable and potentially interpretable by others attending to the robot. 

\begin{figure*}
\centering
\includegraphics[width=0.94\textwidth]{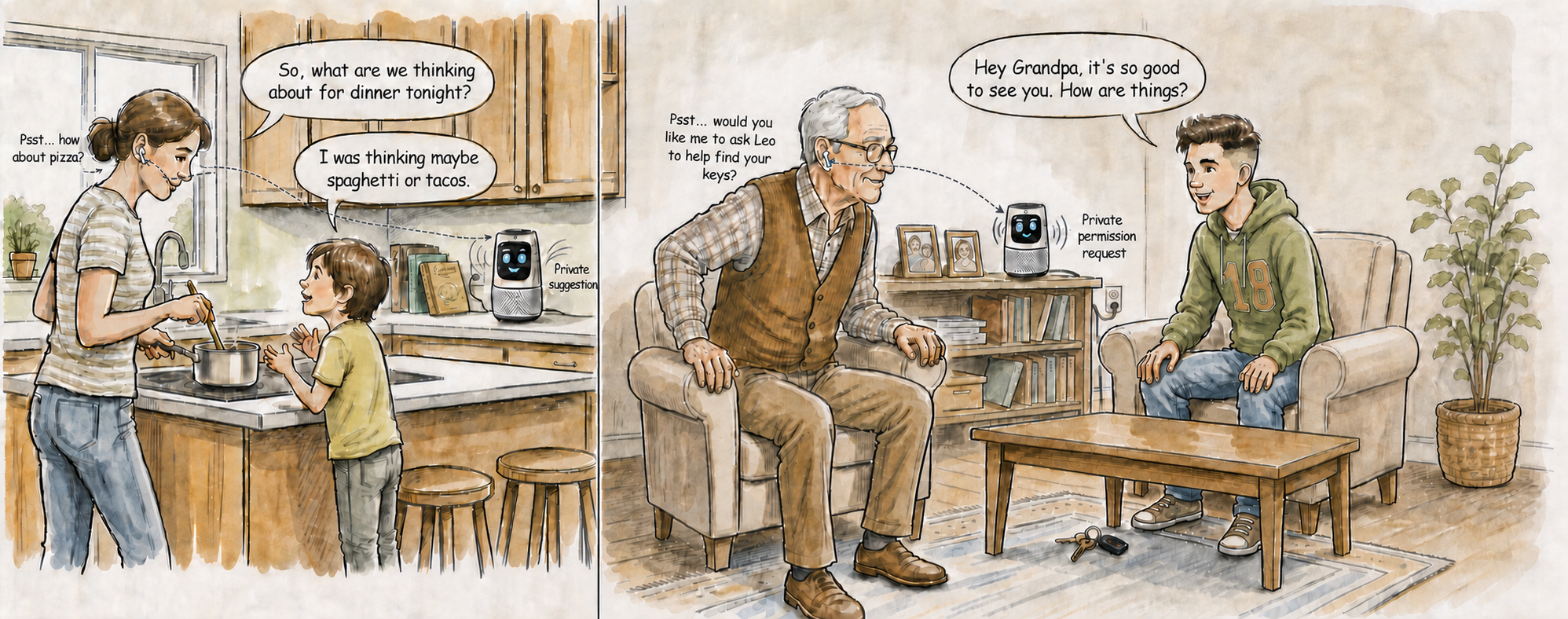}
\Description{Two illustrated hypothetical domestic applications of discreet
interaction with social robots. On the left, a robot sends a suggestion to a mother while she cooks with a child. On the right, a robot offers assistance during a conversation between an older adult and a younger visitor. In both scenes, in-ear audio and a deliberate rightward head-tilt response illustrate a bidirectional exchange embedded within an ongoing shared activity.}
\caption{Two hypothetical scenarios illustrating discreet interaction. On the left, a social robot suggests ordering pizza to a mother via in-ear audio instead of raising the idea openly in front of the child. On the right, the robot uses the wearable-mediated channel to request permission before making a potentially embarrassing suggestion to an older adult. Both scenes illustrate a bidirectional exchange embedded within an ongoing shared activity. Generated with ChatGPT 6.}
\label{fig}
\end{figure*}

In this work, we use \textbf{\textit{discreet interaction}} to
describe a bidirectional exchange between an agent and a human within a shared activity. The exchange
provides differential information access by making message
content available to that person while making their response
less readily interpretable to co-present others. We treat discretion as a graded, observer-dependent design goal; it does not imply invisibility, confidentiality, or guaranteed privacy \cite{Pohl2019SubtleInteraction,Uhde2023BeyondHiding}. Wearable devices provide one possible basis for implementing such an exchange. Figure~\ref{fig} illustrates two hypothetical domestic scenarios in which such an exchange may be embedded within an ongoing shared activity. Ear-worn devices can selectively deliver in-ear audio to an individual while incorporating inertial sensors capable of sensing head movements \cite{Kawsar2018Earables,Min2018Exploring,Li2022EarIO}. 

To investigate such an exchange, we created a bidirectional, wearable-mediated channel combining in-ear audio with a deliberate rightward head-tilt response using eSense\footnote{https://www.esense.io/}.
We evaluated the channel using a screen-based conversational
agent representing a social robot, referred to as the robot
character. Our empirical focus is wearable-mediated interaction
with this conversational agent during a shared activity.
The evaluation required a setting in which exchanged information
could influence subsequent actions while another person
remained an active participant. 

We therefore developed a customized three-player card game inspired by rock--paper--scissors. Prior studies have used familiar competitive games to investigate social responses to a robot's fair or unfair behavior \cite{Short2010NoFair,Litoiu2015Evidence}. Here, the game served a different purpose: participants played in a triadic setting, and the competitive choices gave task-relevant consequences to the information exchanged between the wearer (participant who used the channel) and the robot character. The non-wearer (participant not informed about the channel) was therefore not a passive observer but a co-present participant with opportunities to notice or infer the channel's use.

In-ear audio carried messages from the robot character to the wearer. In the reverse direction, detection or non-detection of a qualifying rightward head tilt from the wearer determined the displayed response state and, consequently, the robot character's next action. Channel-mediated turns were embedded within the game's normal turn structure, and the information exchanged could affect the wearer's or the robot character's subsequent opponent selection. The two turn types assigned the wearer and robot character distinct roles within the prescribed information--response--action sequence, reflecting the procedural interdependence described in human--autonomy teaming research \cite{ONeill2022HumanAutonomyTeaming,Traeger2020VulnerableRobots,Takko2021HumanAgentCoordination}. 

To evaluate the wearable-mediated channel in the shared game, we pose the following research questions:

\begin{itemize} [topsep=6pt, itemsep=5pt]
    \item \textbf{RQ1:} What does the game indicate about the formative, context-bound feasibility of discreet interaction through the wearable-mediated channel?

    \item \textbf{RQ2:} How do wearers perceive pressure/tension when playing the game with versus without discreet interaction through the wearable-mediated channel?

    \item \textbf{RQ3:} How do wearers perceive the wearable-mediated channel's usability after using it for discreet interaction in the game?
\end{itemize}

We use feasibility in a formative, context-bound sense: whether the prescribed exchange could be enacted through the wearable-mediated channel and what wearer and non-wearer accounts reveal about the conditions that supported or constrained it. We treat perceived pressure/tension and immediate usability as separate wearer-reported outcomes. Alongside these research questions, exploratory interviews examined wearers' accounts of channel use, non-wearers' pre-disclosure retrospective accounts of noticing or inference, and both groups' post-disclosure interpretations. We also examined participants' suggestions for alternative interaction techniques, modalities, and application contexts to inform future research on discreet interaction in shared settings.

This study makes three contributions: (I) We describe \textit{discreet interaction} in terms of
bidirectional exchange and differential information access
within a shared ongoing activity. 
(II) We instantiate this concept through a wearable-mediated channel and a controlled three-player card game in which a non-wearer remains an active participant. (III) We provide a formative characterization of protocol consistency, wearers' immediate experience, and wearer and non-wearer interpretations of differential information access via a wearable-mediated channel.

\section{Related Work}

We position our study within research on selective
information access, discreet interaction techniques,
and teamwork and group dynamics involving conversational agents and robots.

\subsection{Selective Information Access}

Social robots use gaze and other nonverbal cues to indicate whom they are addressing and to coordinate participation in conversation. Robot gaze can guide attention \cite{Admoni2017Socialeyegaze}, support inferences about intentions \cite{Mutlu2009NonverbalLI}, and establish participation roles, such as who receives an utterance and who observes the conversation \cite{Mutlu2009FootingIH}. Embodied conversational agents similarly combine gaze, gesture, and speech to manage turn-taking in multi-party interaction \cite{bohus_facilitating_2010}. These cues help distinguish an intended recipient from other participants, but they do not by themselves restrict who can access the conversational content. Selective access to content can also coexist with physical visibility. In a qualitative study of robot interaction in a school setting, Oleson et al.\ \cite{Oleson2022PublicPrivate} reported that teens found comfort in remaining physically visible while their conversations were audibly private. This finding illustrates how an exchange can remain part of a shared social setting while limiting others' access to what is communicated. Building on this, we investigate selective information exchange bidirectionally through a wearable-mediated channel. The design combines messages delivered selectively to the wearer with a response intended to be less readily interpretable to co-present others. We examine this within a shared activity in which the non-wearer remains an active participant.

\subsection{Discreet Interaction Techniques}

Discreet interaction techniques seek to limit how readily others notice or interpret device use. Researchers have explored several approaches to this goal, including concealed interaction, familiar-looking actions, and subtle bodily input. M{\"a}kel{\"a} et al.\ \cite{makela_hidden_2021} examined concealed information acquisition and text entry on smartphones and wearables. \textit{Itchy Nose} \cite{Lee2017Itchynose} detects finger movements on the nose through smart-eyewear sensors, while \textit{ChewIt} \cite{Gallego2019ChewIt} supports hands-free input through an interface inside the mouth. Costanza et al.\ \cite{Costanza2007Intimate} evaluated both the usability and noticeability of motionless gestures detected through muscle activity. These studies highlight two distinct questions for discreet interaction: whether a system can reliably detect an input, and whether nearby people notice or interpret the interaction.

Ear-worn devices provide a platform for investigating these questions by combining sensed input with audio output \cite{Kawsar2018Earables, Roddiger2022SensingEarables}. Prior work demonstrates a range of input possibilities. Earbud motion data have been used to recognize nodding and head shaking \cite{Min2018Exploring, Gashi2021HeadGestures}, and \textit{MorsEar} \cite{Chugh2026MorsEar} uses inertial sensing for covert messaging. Other approaches capture activity beyond head movement: \textit{EarRumble} \cite{Roddiger2021EarRumble} detects pressure changes from voluntary middle-ear muscle contraction, while \textit{EarIO} \cite{Li2022EarIO} tracks facial movements acoustically. Related work on head-worn devices also supports recognition of silently articulated speech \cite{srivastava2024whispering}. These systems establish sensing capabilities, but understanding their potential for discreet interaction also requires examining how the interaction is experienced by wearers and perceived by others.

Research on social acceptability provides a basis for this broader evaluation. Koelle et al.\ \cite{Koelle2020SocialAcceptability} review relevant methods, measures, and design strategies, while empirical studies examine acceptance of head- and eye-based inputs \cite{Hossain2021Exploring}, interpretations of gestures and their effects \cite{Montero2010YouCanTouch}, and differences between performer and observer perspectives \cite{Alallah2018PerformerObserver}. This literature also shows why noticeability alone is insufficient to explain social responses. 

Research on witnesses' experiences considers factors beyond visibility \cite{Uhde2023BeyondHiding}, while work on candid interaction examines the trade-off between making device activity apparent to nearby people and protecting user privacy \cite{ens_candid_2015}. These findings motivate evaluating the wearable-mediated channel from both wearer and non-wearer perspectives, distinguishing technical detection, interaction noticeability, and social acceptability.

\subsection{Teamwork and Group Dynamics}
Research on robots in groups examines how robot behavior influences interactions among human group members \cite{Sebo2020RobotsInGroups}. Studies have explored how robots can address unequal conversational engagement \cite{Matsuyama2015FourParticipant}, balance participation through adaptive gaze \cite{Gillet2021RobotGaze}, and influence human--human conversation and perceptions of the group through expressions of vulnerability \cite{Traeger2020VulnerableRobots}. These studies connect to the broader perspective of interaction-shaping robotics, which examines how robots influence interactions between other group members \cite{gillet_interaction-shaping_2024}. From this perspective, a wearable-mediated exchange warrants attention not only as an interaction with the wearer, but also for its implications for other participants.

These implications may depend on how participants' interests align. Human--autonomy teaming research describes humans and autonomous agents as occupying distinct roles and working interdependently toward a common goal \cite{ONeill2022HumanAutonomyTeaming}. Shared activities, however, can also involve competing interests, and human--agent games provide settings for examining cooperation alongside self-interested choices \cite{Takko2021HumanAgentCoordination}. Selective information access therefore raises questions about how participants understand the robot's role and the exchange when information available to one participant may matter to another in a shared activity. We examine these questions through retrospective interviews with wearers and non-wearers, focusing on how they interpret differential information access within a shared activity in which their interests may compete.

\section{Study Setup}

This section describes the robot character, in-ear wearable, card game, and experimental setup used to evaluate discreet interaction.

\subsection{Robot Character and In-Ear Wearable}

We used a screen-based conversational agent representing a
social robot, rendered as an animated 3D robot character
\cite{Nilgar2024User,Nilgar2025Privacy,wang2021designing}
(see Figure~\ref{fig:Overview_Game}(b)).
The character was torsoless and could hover slightly, slide
laterally, orient toward participants to indicate directed
attention, and display animated facial expressions such as
happiness, sadness, or a thinking state. A researcher controlled
its predefined speech, expressions, and movements through a
Wizard-of-Oz setup (see Section~\ref{GUI}). 

We used the eSense in-ear wearable \cite{Kawsar2018Earables, Min2018Exploring, Odoemelem2019ESenseRobot} for the wearable-mediated channel between the robot character and the wearer. In the discreet condition, the device delivered in-ear audio and captured the deliberate rightward head-tilt response. The eSense inertial measurement unit (IMU) streamed six-axis accelerometer and gyroscope data at 50 Hz. During initialization, the wearer held their head still while the system averaged 250 samples (approximately 5 seconds) to estimate neutral pitch and roll offsets. A complementary filter combined the accelerometer- and gyroscope-derived orientation estimates. The detector classified an offset-corrected roll $\phi$ satisfying $-10^\circ \leq \phi \leq -5^\circ$ as a small rightward head tilt and $\phi < -10^\circ$ as a large rightward head tilt. Either classification constituted detection of a qualifying rightward head tilt and set the displayed response state to ``yes.'' After each binary-response prompt, the researcher monitored the detector output for three seconds. If no qualifying rightward head tilt was detected during this interval, the displayed response state was set to ``no.''

\subsection{Card Game Design}
\begin{figure*}[t]
    \centering
    \begin{minipage}{0.98\textwidth}
        \centering

        \begin{subfigure}[t]{\linewidth}
            \centering
            \includegraphics[width=\linewidth]{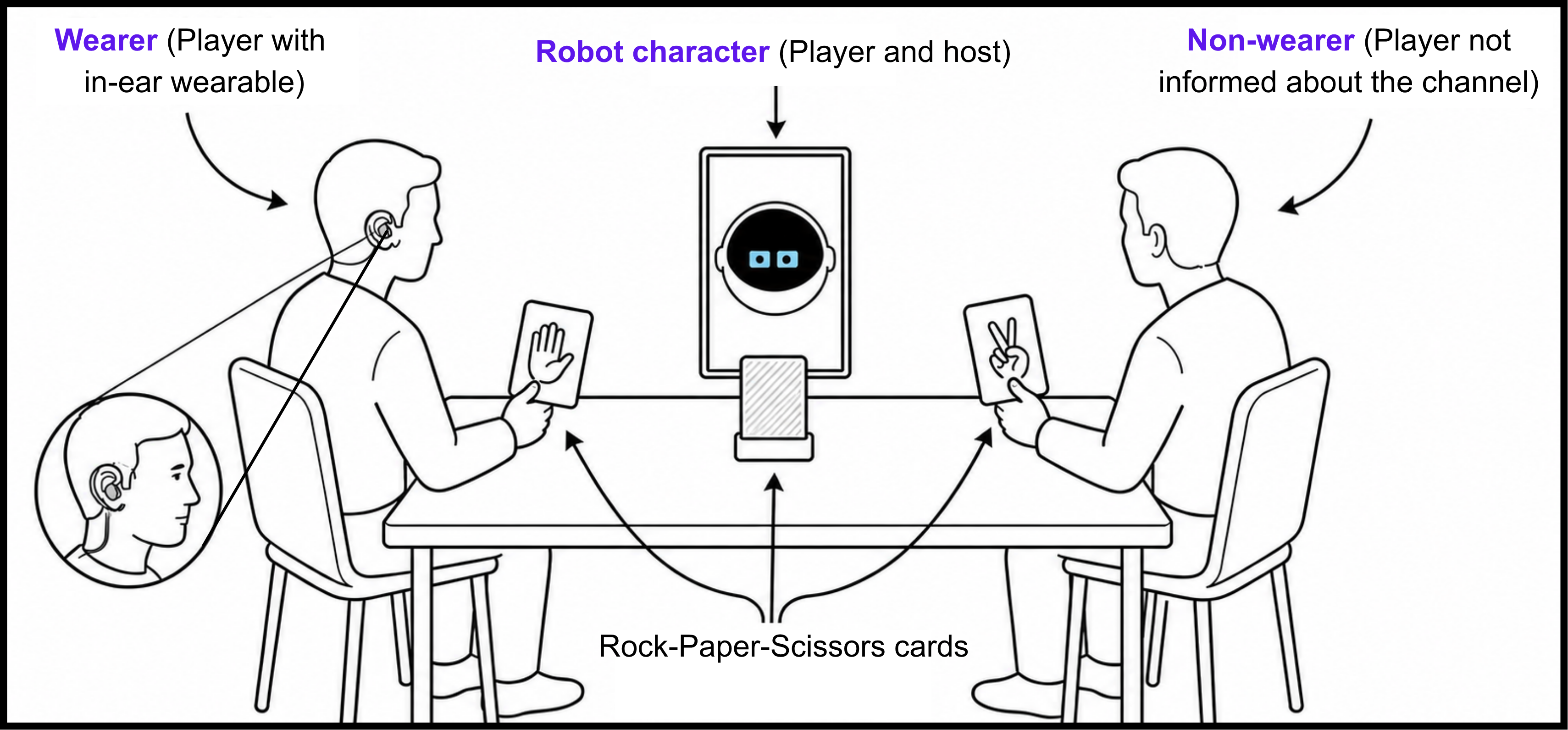}
            \caption{}
            \label{fig:setup-overview}
        \end{subfigure}

        \par\medskip

        \begin{subfigure}[t]{0.22\linewidth}
            \centering
            \includegraphics[width=\linewidth]{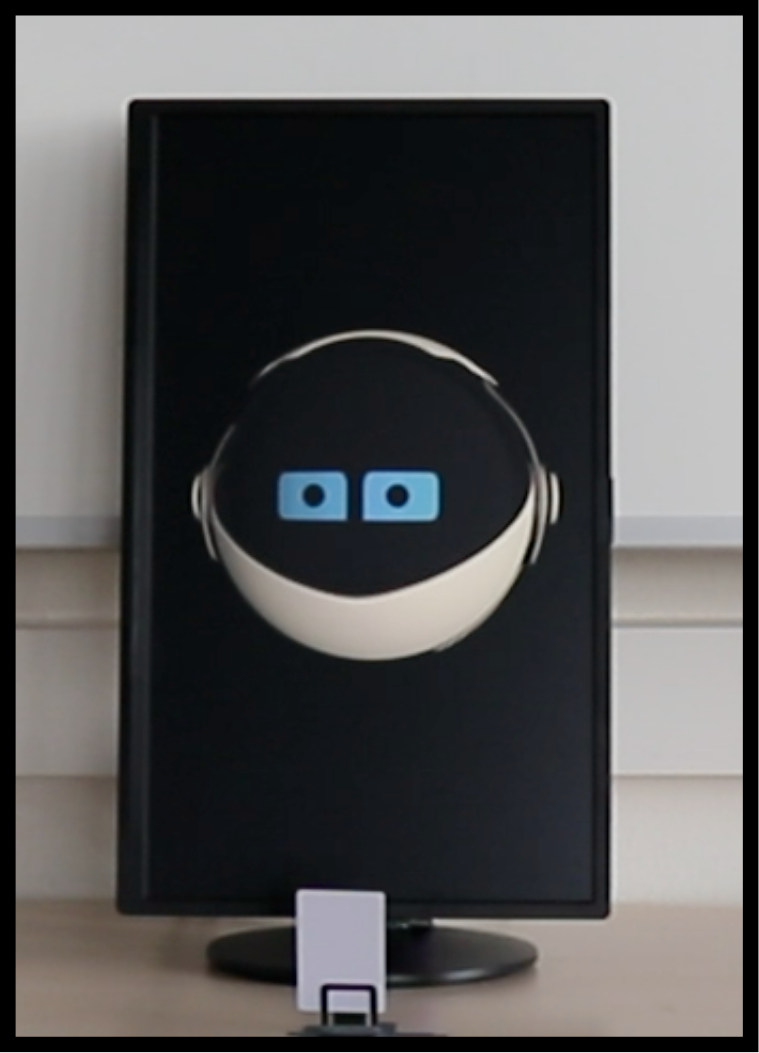}
            \caption{}
            \label{fig:setup-robot}
        \end{subfigure}\hfill
        \begin{subfigure}[t]{0.265\linewidth}
            \centering
            \includegraphics[width=\linewidth]{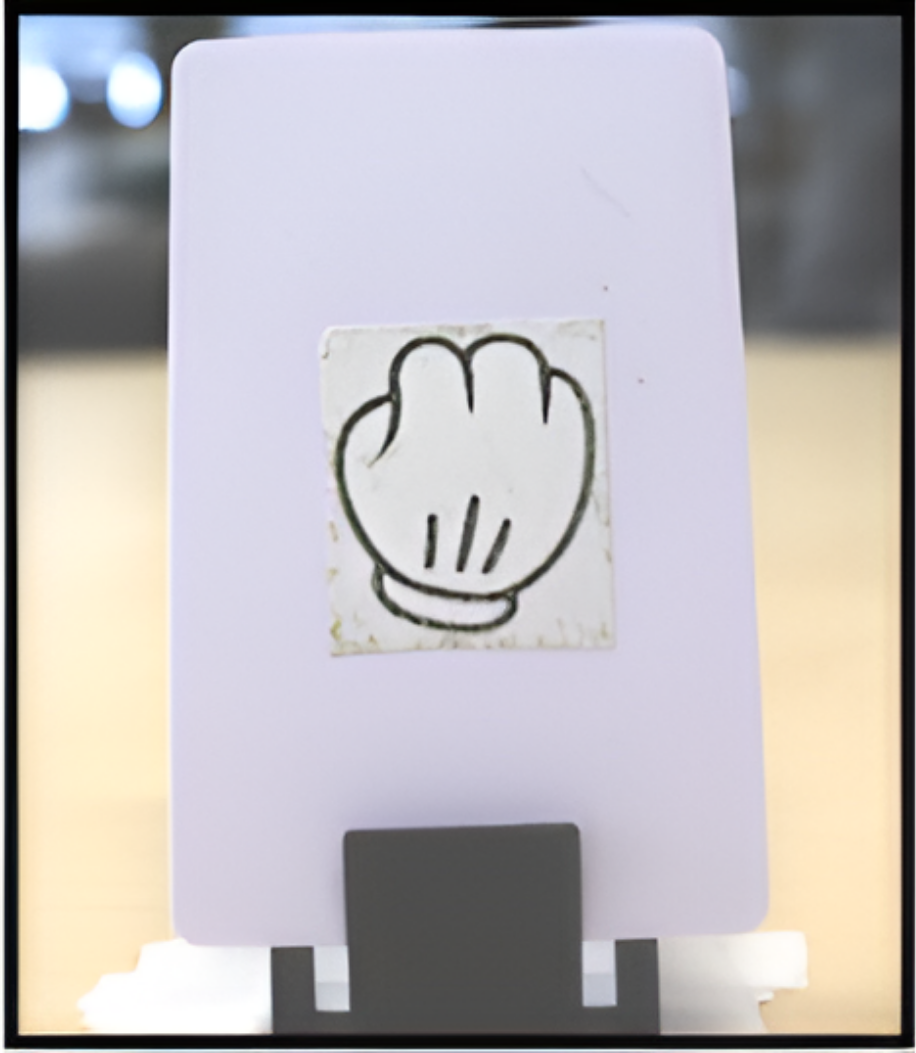}
            \caption{}
            \label{fig:setup-card}
        \end{subfigure}\hfill
        \begin{subfigure}[t]{0.495\linewidth}
            \centering
            \includegraphics[
                origin=c,
                width=\linewidth
            ]{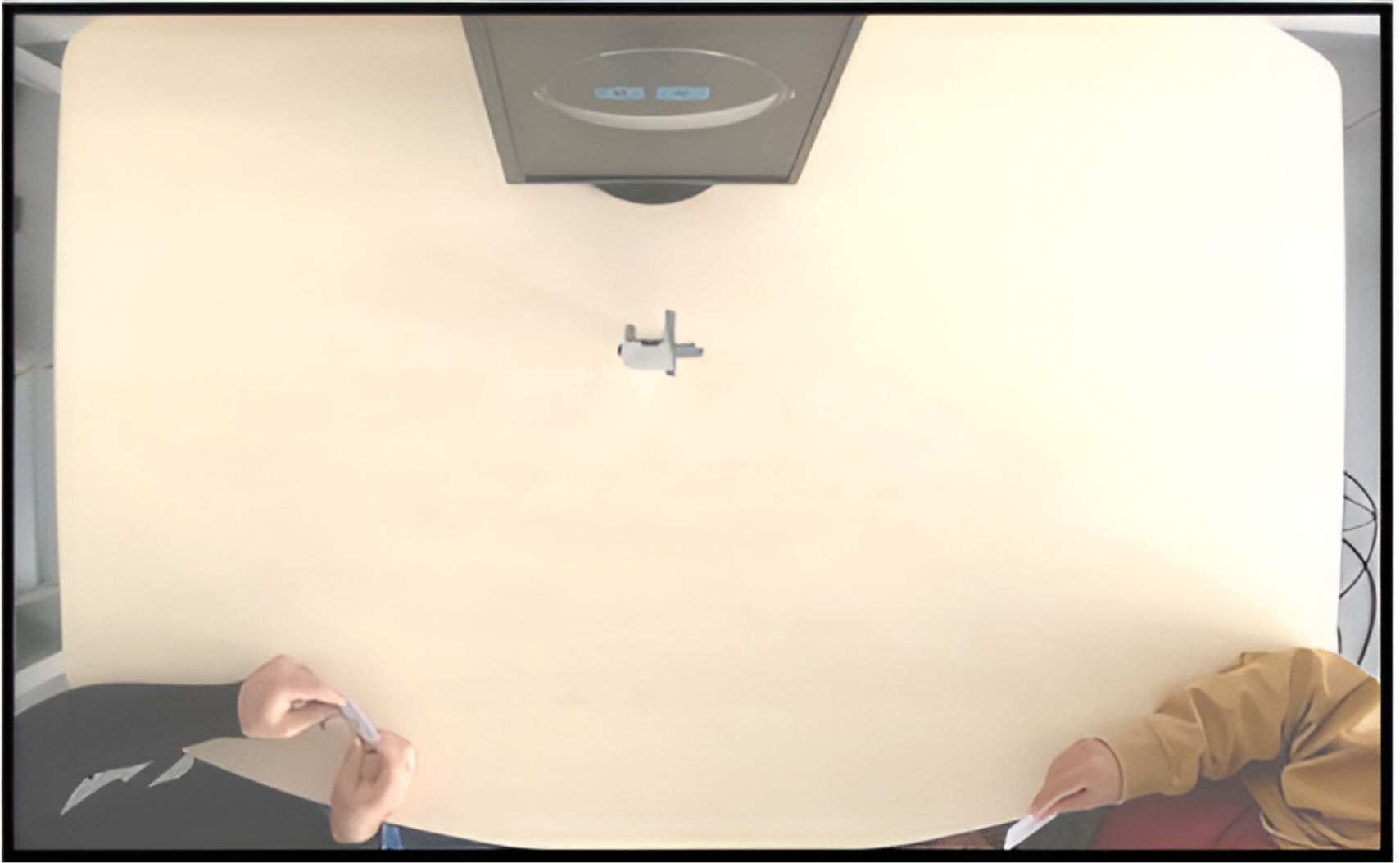}
            \caption{}
            \label{fig:setup-overhead}
        \end{subfigure}
    \end{minipage}

\caption{Study setup for the three-player card game:
(a) schematic showing the wearer and non-wearer seated across
a table, with the screen-based robot character acting as both
host and fellow player. In the discreet condition, the wearer
used an in-ear wearable to receive in-ear audio and perform
a deliberate rightward head-tilt response. The non-wearer
remained an active participant and was not informed about
the wearable-mediated channel until after gameplay;
(b) the robot character displayed on the screen;
(c) the cardholder camera feed showing the robot character's
card; and
(d) the overhead camera feed showing the table and the
participants' revealed cards.
Both camera feeds were available to the researcher
controlling the robot character.}
    \label{fig:Overview_Game}

    \Description{Four panels illustrate the study setup.
    The top schematic shows two participants seated across
    a table with a screen-based robot character between them.
    The wearer uses an in-ear device, while the non-wearer
    is not informed of the channel.
    The bottom panels show the robot character on the screen,
    a rock card in the cardholder camera feed, and an overhead
    camera view of the table and participants holding cards.}
\end{figure*}

For this formative evaluation, we needed consequential information exchange within an activity shared with a co-present non-wearer. Rather than asking participants to judge a video or enact prescribed roles, we used a controlled game in which they directly experienced the action--outcome sequence. We designed a three-player card game inspired by rock--paper--scissors \cite{Short2010NoFair, Litoiu2015Evidence}, using three physical cards---rock, paper, and scissors---as tokens for two participants and the robot character (Figure~\ref{fig:Overview_Game}(a)). The robot character acted as both host and fellow player. 

At the beginning of each round, the researcher activated the robot GUI's random-selection function, and the robot character announced which participant would shuffle and deal the cards. The robot character's card was placed in the cardholder (Figure~\ref{fig:Overview_Game}(c)), and the two participants privately inspected their own cards. The robot GUI also generated a randomized order in which each player served as the guesser once over the three rounds; the robot character announced the designated guesser at the start of each round.

The guesser chose one of the other two players, aiming to select the player whose card the guesser's card could defeat. The guesser and chosen opponent then revealed their cards simultaneously. The standard rules determined the round winner: rock defeats scissors, scissors defeats paper, and paper defeats rock. If the opponent held the weaker card, the guesser won the round; otherwise, the opponent won. After three rounds, the player with the most round wins was the overall winner; games without a unique winner were recorded as ties.

\subsection{Game Setup and Character Control}\label{GUI}
Figure~\ref{fig:Overview_Game}(a) shows the participants' seating arrangement. One participant received the in-ear wearable before the discreet condition. The two participants sat across a table, with the device worn on the side facing away from the other participant. This arrangement made the device less directly visible but did not guarantee that the device or its function would go unnoticed. The researcher used two camera feeds to monitor the game scene. The cardholder camera was mounted below the screen displaying the robot character to capture its card when it was placed in the cardholder (Figure~\ref{fig:Overview_Game}(c)). The overhead camera captured both human players' revealed cards on the table to determine the outcomes (Figure~\ref{fig:Overview_Game}(d)).

We used a Wizard-of-Oz setup~\cite{Dow2005WizardOfOz} in which
a researcher remotely controlled the robot character, allowing
us to examine discreet interaction without uncertainty from
autonomous perception and decision-making. The researcher operated the robot character and the wearable-mediated channel through two graphical user interfaces (GUIs).
The robot GUI used preconfigured buttons to control the robot character's publicly observable behavior, including speech, facial expressions, and movements such as orienting toward a participant. Its speech output consisted of predefined phrases for card shuffling, turns, winners, and random guesses. 

\begin{figure*}[t]
    \centering
    \includegraphics[width=\textwidth]{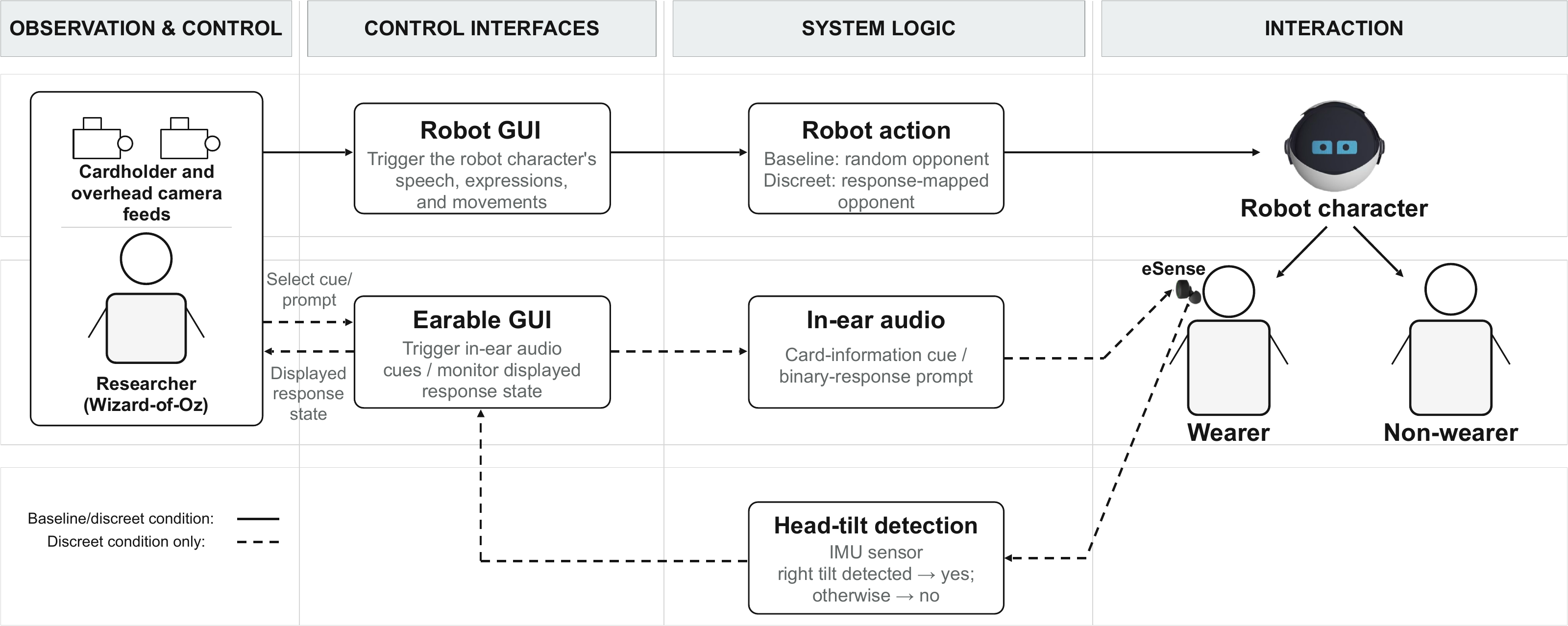}
    \Description{Interaction-flow diagram arranged from observation and researcher control to the robot character and participants. Solid arrows show the flow used in the baseline and discreet conditions: the researcher observes the cardholder camera feed and the overhead camera feed, operates the robot GUI, and triggers the robot character's publicly observable behavior. Dashed arrows show the additional discreet condition flow: the researcher selects a card-information cue or binary-response prompt in the earable GUI; the wearer receives in-ear audio and can perform the deliberate rightward head-tilt response; the eSense IMU supports detection or non-detection of a qualifying rightward head tilt, and the displayed response state informs the researcher.}
\caption{Information flow in the Wizard-of-Oz setup.
The researcher monitored the cardholder and overhead camera
feeds and controlled the screen-based robot character's
speech, expressions, and movements through the robot GUI.
In the discreet condition, the researcher used the earable
GUI to deliver a card-information cue when the wearer was
the guesser or a binary-response prompt when the robot
character was the guesser. Following the binary-response
prompt, detection of a qualifying rightward head tilt set
the displayed response state to ``yes,'' while non-detection
within three seconds set it to ``no.''
The researcher monitored this state to guide the robot
character's opponent selection. In the baseline condition,
the robot GUI selected the robot character's opponent
using its random-selection function.}
    \label{fig:interaction_flow}
\end{figure*}

The earable GUI for the wearable-mediated channel allowed the researcher to trigger prerecorded card-information cues when the wearer was the guesser, issue a binary-response prompt when the robot character was the guesser, and monitor the displayed response state in real time. Figure~\ref{fig:interaction_flow} summarizes the information flow during the game.

\section{Study Design}

\subsection{Recruitment}
For this initial formative evaluation, we recruited throughout the planned data-collection period rather than using a power-based stopping rule. We recruited 20 pairs (40 participants), most of whom were university staff or students. Participation was voluntary and uncompensated. Prior familiarity between pair members was not recorded;
therefore, we cannot characterize the pairs as strangers
or acquaintances. Among the wearers, 17 identified as male and 3 as female; among the non-wearers, 15 identified as male and 5 as female. Wearers were between 22 and 36 years of age ($M = 27$, $SD = 3.50$, $Mdn = 26.5$), and non-wearers were between 23 and 30 years of age ($M = 26.3$, $SD = 2.31$, $Mdn = 26$). Nineteen wearers and all 20 non-wearers reported holding a degree, while one wearer reported holding a high school degree. 

Among the wearers, 17 reported Germany and three reported India
as their country of residence. Among the non-wearers, 18 reported
Germany, one reported India, and one reported Italy. All participants reported no hearing impairments.
Wearers reported using headphones frequently ($n = 11$), sometimes ($n = 6$), or rarely ($n = 3$). Non-wearers reported using headphones frequently ($n = 15$), sometimes ($n = 4$), or rarely ($n = 1$). Six wearers and five non-wearers reported prior experience interacting with conversational or social robots.

\subsection{Experimental Design}
We used a within-pair design: each pair completed both the baseline and discreet conditions, and the wearer and non-wearer assignments remained fixed. We balanced condition order across pairs using an alternating allocation sequence: 10 pairs completed the baseline condition first, and 10 completed the discreet condition first. Both conditions consisted of three rounds. In each condition, the wearer, non-wearer, and robot character each served once as the guesser, keeping the number of turns balanced across players.

\textbf{Baseline condition.}
In the baseline condition, the wearer received no card-information cue and selected an opponent based on their own judgment. When the robot character was the guesser, the researcher activated the random-selection function in the robot GUI, which programmatically selected an opponent.

\textbf{Discreet condition.}
The wearable-mediated channel was used when either the wearer or the robot character was the guesser. When the wearer was the guesser, the researcher played a prerecorded card-information cue (e.g., ``I have the rock card'') after both participants had privately viewed their cards and before the wearer chose an opponent. The card-information cue identified the robot character's card. Because all three players held different cards, the wearer could combine this information with their own card to infer the non-wearer's card. The wearer could choose either opponent, but the prescribed coordination protocol specified choosing the opponent whose card the wearer's card could defeat. When the robot character was the guesser, it issued a binary-response prompt based on its card. 

The prompt asked whether the wearer held the card that the robot character's card could defeat. The card-implied response was ``yes'' if the wearer held that card and ``no'' otherwise. For example, if the robot character held rock, the prompt was ``If you have the scissors card, please tilt your head to the right.'' The researcher monitored the detector output for three seconds after the binary-response prompt. Detection of a qualifying rightward head tilt set the displayed response state to ``yes,'' while non-detection set it to ``no.'' The robot character then selected the wearer as its opponent for ``yes'' and the non-wearer for ``no.'' The wearable-mediated exchange was embedded without changing the game's turn structure. On their guesser turn, the non-wearer could freely choose either opponent. 

The non-wearer was not told about the wearable-mediated channel until after the game. Together, the two turn types implemented both directions of the interaction: the robot character delivered information before the wearer chose an opponent, and detection or non-detection of a qualifying rightward head tilt set the displayed response state used for the robot character's subsequent choice.

\subsection{Procedure}
Participants first completed a questionnaire about demographics and technology experience using two computers, one in the main laboratory and one in an adjacent room. The participant who used the main-laboratory computer wore the in-ear wearable in the discreet condition. We use \emph{wearer} and \emph{non-wearer} only as reporting labels, not as roles presented to participants. Before the first condition, the researcher explained the card-game rules, three-round structure, and guesser's responsibilities. Participants were told that the robot character would act as both host and fellow player. Participants were free to talk to one another during gameplay.
They were required to keep their cards concealed until the game
called for them to be revealed. The content of conversations
between participants during gameplay was not recorded. Before the discreet condition, while the non-wearer waited in a separate room, the researcher explained the in-ear audio and deliberate rightward head-tilt response to the wearer. The wearer was not informed that the researcher operated the wearable-mediated channel; during gameplay, the audio and response were presented as communication with the robot character through the channel. 

The researcher then calibrated the in-ear wearable and asked the wearer to perform the response several times to confirm correct detection. The audio volume was fixed at 50\% for all wearers to standardize delivery and limit potential acoustic leakage, and each wearer confirmed that they could hear it clearly. To maintain a consistent spatial arrangement across pairs, the wearer sat on the left side of the table and the non-wearer on the right (Figure~\ref{fig:Overview_Game}(a)). Each pair completed the baseline and discreet conditions in the order described above. Only wearers completed the perceived pressure/tension and usability questionnaires (see Section~\ref{measures}). 
The game lasted approximately 20 minutes and was followed by a joint semi-structured interview with both members of the pair, lasting approximately 10 minutes. The game, questionnaires, and interviews were administered in English. The study was approved by the Ethics Council of the University of
Siegen (reference: LS\_ER\_29\_2025).

\subsection{Measures}\label{measures}

\subsubsection{Protocol Consistency}
We assessed whether each of the 40 channel-mediated turns followed the prescribed coordination protocol: 20 wearer-guesser turns and 20 robot-character-guesser turns. A wearer-guesser turn was classified as protocol-consistent when the card-information cue correctly stated the robot character's card and the wearer selected the opponent specified by the coordination protocol. A robot-character-guesser turn was classified as protocol-consistent when the binary-response prompt followed the prescribed card mapping, the displayed response state matched the card-implied response, and the robot character selected the opponent indicated by the displayed response state. We used the number of protocol-consistent turns as the measure of protocol consistency. This turn-level measure characterizes procedural enactment of the prescribed exchange; it does not assess non-wearers' noticing or inference, participants' interpretations, or component-level accuracy.

\subsubsection{Perceived Pressure/Tension}
We measured wearers' perceived pressure/tension with the five-item intrinsic motivation inventory (IMI) pressure/tension subscale to assess differences between conditions \cite{McAuley1989Psychometric}. Wearers completed the subscale after each condition using a 7-point response scale from ``Not at All True'' (1) to ``Very True'' (7). Following the standard IMI scoring procedure, we reverse-coded the applicable items and averaged the five responses to calculate one perceived pressure/tension score for each wearer in each condition. All questionnaire responses were complete, and higher scores indicate greater perceived pressure/tension.

\subsubsection{Usability}
We assessed wearers' immediate usability ratings of the wearable-mediated channel with the 30-item USE questionnaire \cite{Lund2001Measuring}, comprising usefulness (8 items), ease of use (11 items), ease of learning (4 items), and satisfaction (7 items). Wearers completed the questionnaire after the discreet condition using a 7-point response scale from ``Strongly Disagree'' (1) to ``Strongly Agree'' (7). For each wearer, we averaged the item responses within each of the four USE dimensions. All USE questionnaire responses were complete.

\subsection{Post-Study Interviews}

We conducted semi-structured interviews after both conditions to explore how participants understood and evaluated discreet interaction. Without mentioning the wearable-mediated channel, the researcher first asked non-wearers whether they had noticed unusual behavior from the robot character or wearer during gameplay. Non-wearers could volunteer any inference about an undisclosed interaction in response. During these questions, the wearer was asked not to reveal or discuss the channel until the non-wearer had completed their pre-disclosure responses. The researcher then disclosed the channel. Subsequent questions invited wearers to discuss the effort required by the head-tilt response, any unease or concern about being noticed, and their overall experience of the channel. Participants in both roles were asked about their interpretations, appropriate application contexts, and possible alternative modalities or techniques. The session concluded with a full debriefing explaining the study's aims.

\subsection{Analysis}

We compared wearers' perceived pressure/tension scores between the baseline and discreet conditions using a two-sided Wilcoxon signed-rank test. The significance threshold was $\alpha = 0.05$. We summarized the USE dimension scores descriptively and calculated two-sided 95\% confidence intervals around the dimension means using Student's $t$ distribution with 19 degrees of freedom.

We conducted an exploratory inductive thematic analysis informed by Braun and Clarke \cite{Braun2006Usingthematic} to identify how participants described their experience of discreet interaction. The interviews were audio-recorded and transcribed verbatim. We used MAXQDA software to organize the transcripts, manage the coding, and retrieve coded excerpts. 

The first researcher examined participants' utterances line by line and assigned descriptive codes derived from their substantive content, including noticing or inference, feeling cheated, usability, potential applications, relationships, and alternative interaction techniques. The second researcher reviewed and refined the coding across the transcripts. The researchers then compared their interpretations through iterative discussion, refined the code definitions, and grouped related codes into higher-order themes. They subsequently checked the themes against the coded excerpts and collaboratively finalized their scope. Participant quotations are identified by role and pair number (e.g., W1 for the wearer and NW1 for the non-wearer in Pair 1).

\section{Results}

\subsection{Protocol Consistency}
Overall, 39 of 40 channel-mediated turns were protocol-consistent. All 20 wearer-guesser turns were protocol-consistent: the card-information cue correctly stated the robot character's card, and the wearer selected the opponent prescribed by the coordination protocol. Nineteen of the 20 robot-character-guesser turns were protocol-consistent. In all 20 robot-character-guesser turns, the binary-response prompt followed the prescribed card mapping, and the robot character selected the opponent indicated by the displayed response state. In the remaining turn, the card-implied response was ``yes.'' Still, no qualifying rightward head tilt was detected, producing a displayed response state of ``no.'' W17 later attributed this outcome to intentionally withholding the protocol-consistent response to make the robot character lose. The interview account therefore indicates deliberate non-cooperation, although the logged non-detection alone cannot distinguish intentional nonresponse from missed detection. This turn remained included in the 40-turn total and accounted
for the single protocol-inconsistent turn.

\begin{figure*}[t]
    \centering

    \begin{subfigure}[t]{0.49\linewidth}
        \centering
        \includegraphics[width=\linewidth]{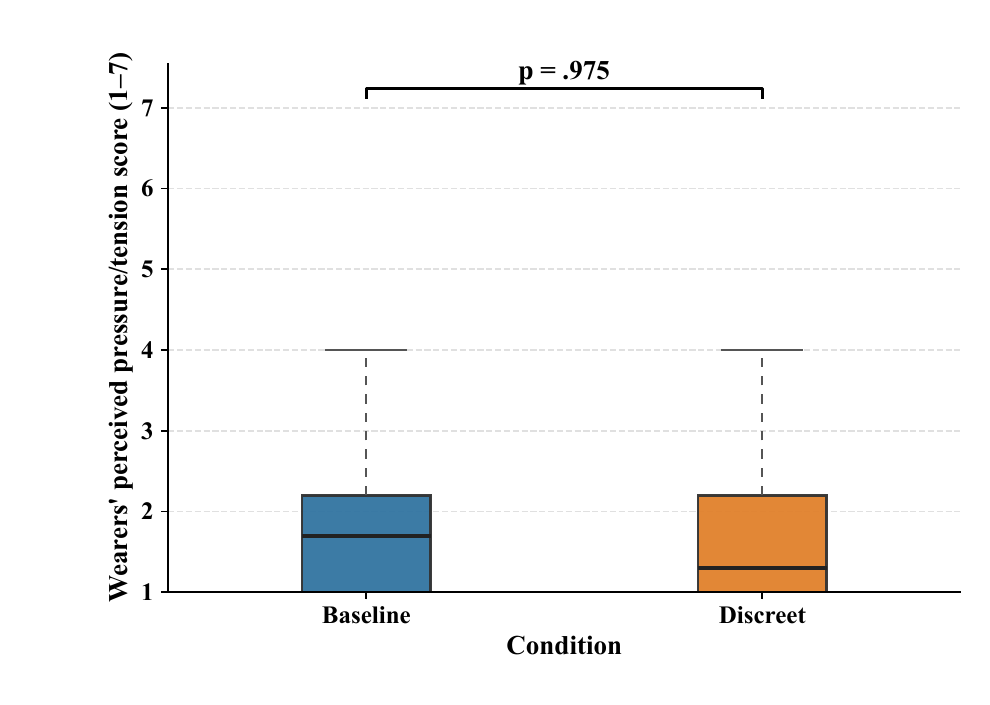}
        \subcaption{Wearers' pressure/tension scores after the baseline
and discreet conditions.
Lower scores indicate lower perceived pressure/tension.}
        \label{fig:pressure}
    \end{subfigure}
    \hfill
    \begin{subfigure}[t]{0.49\linewidth}
        \centering
        \includegraphics[width=\linewidth]{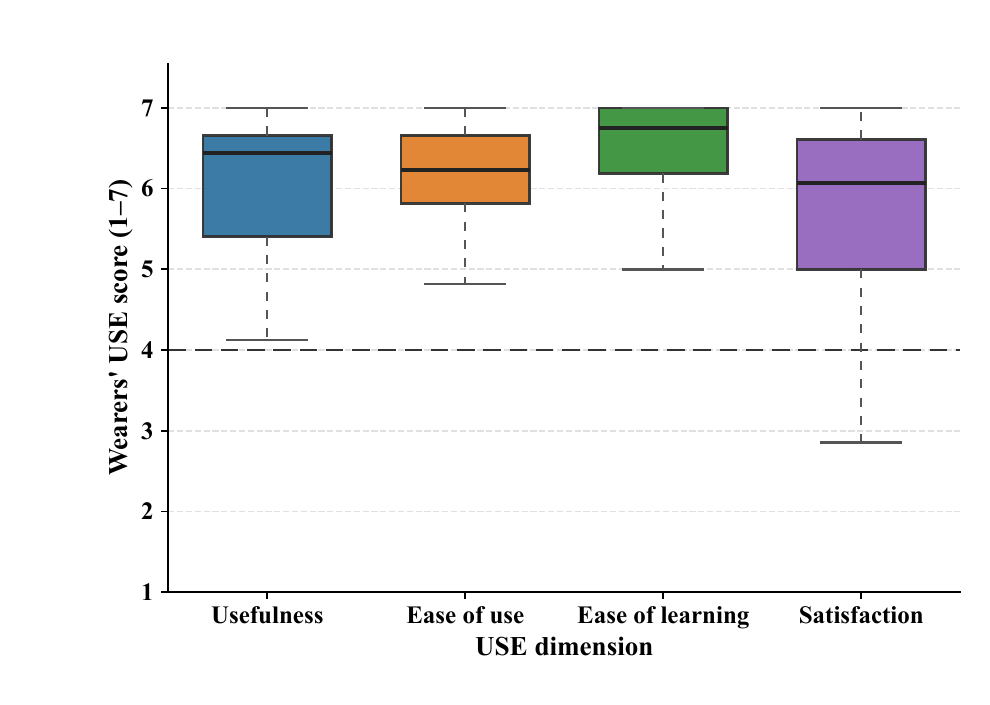}
        \subcaption{Wearers' USE dimension scores after the discreet condition; the reference line marks the scale midpoint (4).}
        \label{fig:usability}
    \end{subfigure}

    \Description{Two panels showing wearer-reported outcomes. Left: boxplots
    of perceived pressure/tension scores in the baseline and discreet
    conditions. Right: boxplots of USE dimension scores for usefulness,
    ease of use, ease of learning, and satisfaction after the discreet
    condition, with the scale midpoint marked.}

    \caption{Wearer-reported perceived pressure/tension and usability outcomes.}
    \label{fig:wearer_results}
\end{figure*}

\subsection{Wearers' Perceived Pressure/Tension}

Wearers reported low perceived pressure/tension in both the baseline
($M = 1.80$, $SD = 0.83$, $Mdn = 1.70$,
$Q_1$--$Q_3 = [1.00, 2.20]$)
and discreet conditions
($M = 1.71$, $SD = 0.91$, $Mdn = 1.30$,
$Q_1$--$Q_3 = [1.00, 2.20]$;
Figure~\ref{fig:pressure}).
Here and below, $Q_1$--$Q_3$ denotes the interval from the
first to the third quartile.
A two-sided Wilcoxon signed-rank test did not detect a
statistically significant difference between conditions
($V = 53.5$, $p = .975$, $r = 0.01$).

\subsection{Immediate Usability}\label{usabilitysection}
Wearers perceived the wearable-mediated channel as useful
($M = 6.06$, $SD = 0.91$, $Mdn = 6.44$,
$Q_1$--$Q_3 = [5.41, 6.66]$, 95\% CI $[5.63, 6.48]$)
and easy to use
($M = 6.20$, $SD = 0.59$, $Mdn = 6.23$,
$Q_1$--$Q_3 = [5.82, 6.66]$, 95\% CI $[5.92, 6.47]$).
These ratings indicate favorable immediate perceptions of
the channel's usefulness and ease of use within the game
(Figure~\ref{fig:usability}). Ease of learning received the highest mean rating
($M = 6.51$, $SD = 0.58$, $Mdn = 6.75$,
$Q_1$--$Q_3 = [6.19, 7.00]$, 95\% CI $[6.24, 6.78]$).
Wearers also rated their satisfaction positively
($M = 5.80$, $SD = 1.06$, $Mdn = 6.07$,
$Q_1$--$Q_3 = [5.00, 6.61]$, 95\% CI $[5.31, 6.29]$).
All four dimension
means exceeded the scale midpoint of 4, showing
favorable initial perceptions of the channel.

\subsection{Exploratory Interview Findings}\label{QF}

We organize the accounts from the joint interviews into five descriptive themes.

\textbf{(I) Pre-Disclosure Attention, Noticing, and Inference:} Non-wearers' accounts distinguished between attending to the
device, interpreting its purpose, and noticing changes in
gameplay. Some reported no awareness of an unusual interaction.

NW5 stated, \textit{``I didn't notice it at all.''}
NW17 explained, \textit{``I never checked the other player's ear.''} These accounts suggest that participants did not necessarily
direct their attention toward the wearable during the game. Seeing the device would not necessarily have revealed its
function. NW9 reflected, \textit{``Even if I had seen the in-ear
wearable, I probably would not have guessed.''}
By contrast, NW14 noticed a change in gameplay, observing that
\textit{``the other player was suddenly winning more rounds.''}
This account illustrates how the consequences of information
exchange could attract attention without establishing awareness
of the channel itself.

\textbf{(II) Post-Disclosure Fairness Interpretations:}
W1 acknowledged the differential information access between the wearer and the non-wearer. Still, W1 downplayed it because the interaction was framed as a game: \textit{``It was a little unfair for the other player, but \ldots\ just a game, that is fine.''} In contrast, NW3 reported feeling cheated after learning about the use of the wearable-mediated channel: \textit{``Afterwards, I feel severely cheated.''} These accounts illustrate contrasting post-disclosure interpretations of the strategic advantage created by the discreet interaction.

\textbf{(III) Perceived Social Role and Selective Cooperation:}
Wearers described the robot character as a partner or teammate. W1 stated, \textit{``I feel like we were playing in a team,''} and W18 said that the robot character felt \textit{``like a part of my team.''} W9 described the robot character as friendly, noting that \textit{``the robot looked into your eyes.''} 

Participants also remarked on the robot character's human-like qualities. For example, NW1 stated, \textit{``The robot felt very human like \ldots\ did not make the person nervous,''} while NW8 described it as \textit{``much more like human type''} because of \textit{``eye contact''} and said that it \textit{``felt social.''} These responses suggest that participants interpreted the discreet interaction in light of the robot character's perceived social role. At the same time, not all wearers acted collaboratively. 

W17 stated, \textit{``I chose not to cooperate with the robot,''} and reported using the robot character's card-information cue to improve their own chances of winning while intentionally withholding the protocol-consistent response to make the robot character lose.

\textbf{(IV) Ease of the Head-Tilt Response and Suggested Alternatives:}
Wearers described the deliberate rightward head-tilt response as easy and low-effort. W1 described it as \textit{``easy \ldots\ just a slight head tilt,''} and W13 said that it was \textit{``fairly easy to use the head gestures.''} 

W12 described initial anxiety before using the response but found it easy in practice: \textit{``At first I felt anxious \ldots\, but when I used it, it was easy.''} Participants also proposed subtle alternative input methods that might be more suitable in other contexts. W7 suggested an \textit{``eyebrow gesture''} as \textit{``more secretive,''} W5 mentioned \textit{``lip reading,''} W12 suggested that a \textit{``blink could work,''} and NW3 proposed \textit{``tapping finger''} or a small \textit{``hand movement.''}

\textbf{(V) Participant-Suggested Application Contexts:}
Some participants contrasted the competitive game with hypothetical contexts in which they considered discreet interaction acceptable or beneficial. NW1 imagined a \textit{``doctor and robot having secret communication''} to \textit{``avoid alarming patients,''} and NW6 described the system as an \textit{``aiding device for nurses''} that could provide reminders of patient information. W18 suggested that it could be used as a \textit{``safety gadget''} in emergencies. 

Participants also mentioned workplace, professional, and smart-home contexts for selective information sharing. NW8 referred to \textit{``non-disclosure meetings''} where the system could help people raise complaints anonymously; NW16 suggested that in-ear audio could help \textit{``while doing a presentation,''} W9 mentioned \textit{``smart-home notifications without alarming guests,''} and W16 described \textit{``private reminders''} for daily task support.

\section{Discussion}

This study evaluated a bidirectional wearable-mediated channel
for discreet interaction using a screen-based conversational
agent representing a social robot in a controlled three-player
card game inspired by rock--paper--scissors. We first discuss protocol consistency and exploratory interview
findings in relation to RQ1, then consider wearers' perceived
pressure/tension and immediate usability in relation to RQ2
and RQ3. We conclude with design implications and limitations.

\subsection{Formative Feasibility Through the Wearable-Mediated Channel (RQ1)}

To address RQ1, we treated feasibility as a formative judgment informed by multiple sources rather than as a single measured outcome. Protocol consistency provided evidence about whether the prescribed exchange could be enacted through the wearable-mediated channel, while the interview themes provided evidence about participants' experiences, noticing, and inference. \textbf{The observed protocol consistency shows that the prescribed exchange could be carried out within the ongoing shared activity.} The two channel-mediated turn types supported communication in both directions without altering the game's normal turn structure. This demonstrated bidirectional capacity at the protocol level. The procedural interdependence resembles human--autonomy teaming accounts in which humans and agents contribute distinct actions \cite{ONeill2022HumanAutonomyTeaming}. 

The interview accounts identify considerations for feasibility beyond procedural enactment. \textbf{Non-wearers differed in what they reported noticing
or inferring about the interaction.} Non-wearer participants reported either not noticing at all (NW5) or not attending to the wearer's ear (NW17). One participant (NW14) noticed a change in the wearer's performance. This illustrates that discretion depends not only on the technical channel but also on observable behavioral consequences. Therefore, discreet interaction does not aim to guarantee undetectability \cite{Pohl2019SubtleInteraction,Uhde2023BeyondHiding}.

W17's account further illustrates why protocol consistency should not be equated with reciprocal cooperation or shared goals. W17 reported using the robot character's card-information cue for personal advantage while intentionally withholding the
protocol-consistent response to make the robot character lose. 
\textbf{Bidirectionality therefore describes the channel's capacity for communication in both directions, not an obligation for the wearer to reciprocate.} This selective use is consistent with prior work showing that participants in human--agent games may balance coordination with self-interest \cite{Takko2021HumanAgentCoordination} and shows why participant accounts were needed to explain departures from the protocol.

\subsection{Wearers' Experience (RQ2 and RQ3)}

\textbf{Wearers reported low perceived pressure/tension in both conditions.} Scores were near the lower end of the 1--7 scale, and the two-sided Wilcoxon signed-rank test did not detect a statistically significant difference between the baseline and discreet conditions. Because the scores were close to the scale floor, this result should not be interpreted as evidence that the conditions were equivalent. Within these limits, wearers reported low pressure/tension during the brief game.

\textbf{Wearers' immediate usability ratings were favorable.} Wearer accounts described the deliberate rightward head-tilt response as easy and low-effort (W1, W13), while W12 reported initial anxiety but found the response easy in practice. These accounts complement prior demonstrations of head-gesture sensing through ear-worn devices by describing how wearers experienced the gesture within an integrated channel \cite{Min2018Exploring,Gashi2021HeadGestures}. Because the channel also provided strategically useful information during the game, usefulness ratings may partly reflect that advantage. These findings describe wearers' immediate experience after brief use and do not establish long-term usability, comfort, acceptance, or use.

\subsection{Social Interpretations of the Channel}
Some \textbf{wearers described the discreet interaction as teamwork
with the robot character} (W1, W18). However, one wearer also
acknowledged that \textbf{the information advantage could be unfair
to the non-wearer} (W1). After learning about the channel,
one non-wearer reported feeling cheated (NW3).
These accounts suggest that the same channel may be viewed as supportive when it enables collaboration and as unfair when it gives one participant an information advantage in a competitive decision-making setting. 

Participants suggested potential applications such as healthcare and nursing support (NW1, NW6), emergency assistance (W18), presentations (NW16), and smart-home notifications (W9), where selectively delivered information might be beneficial or less disruptive. Social acceptability is therefore not a property of the communication technique alone; it is shaped by the scenario, participants' goals, and the distribution of benefits within the group \cite{Sebo2020RobotsInGroups,Traeger2020VulnerableRobots,gillet_interaction-shaping_2024}. In this study, the reactions also concerned competitive advantage and subsequent disclosure, so they should not be attributed to the channel alone but to the interaction as a whole.

\subsection{Design Implications}
\textbf{Discreet interaction should be designed for all participants affected by the exchange.} Mean ratings for all four USE dimensions exceeded the scale midpoint, but these ratings do not establish the channel's appropriateness for non-wearers. Participants described the same arrangement in terms of teamwork, unfairness, and feeling cheated, suggesting that perceived appropriateness may depend on the exchange's purpose and consequences. Designers should therefore consider whose interests the channel serves, how it affects relationships within the group, and what co-present others should know about it. Awareness of its existence or general function may be sufficient in some settings, while others may call for consent or further disclosure, even when message content remains selectively accessible \cite{Koelle2020SocialAcceptability,Alallah2018PerformerObserver,ens_candid_2015}.

Alongside these considerations for non-wearers, the design should protect wearer agency. One wearer accepted information from the robot character but intentionally withheld the protocol-consistent response (W17), illustrating that access to the channel does not create an obligation to cooperate. However, the current system mapped non-detection of a qualifying rightward head tilt to the displayed response state ``no,'' conflating an intentional negative response with refusal, missed detection, or no registered input. Future designs should distinguish these states and allow wearers to confirm or correct a response before the robot character acts on it.

Supporting wearer agency also involves providing suitable ways to respond. Participants suggested eyebrow gestures, blinks, and finger taps, motivating investigation of personalized response techniques suited to wearers' comfort, abilities, and ongoing activities. These alternatives were not evaluated here, and personalization alone does not establish that an exchange will go unnoticed or be socially acceptable. Similarly, participant-suggested applications, including healthcare scenarios, require separate evaluation of their suitability for both wearers and affected non-wearers. \textbf{We do not endorse the use of this design to facilitate cheating, deception, or unfair advantage.} We intend this work to support the responsible exploration of discreet interaction.

\subsection{Limitations and Future Work}
The study involved 20 pairs drawn from a predominantly male, highly educated university population. Each participant retained the same wearer or non-wearer role across both conditions, seating positions were fixed, and prior familiarity between pair members was not recorded. Although the brief competitive game gave the exchanged information immediate task consequences, the setting limits the transferability of the findings to longer or non-competitive interactions. Future studies should recruit more diverse populations, vary who wears the device, and examine these additional interaction settings. Protocol consistency does not establish component-level accuracy or autonomous-system performance. It provides context-bound evidence rather than establishing long-term or real-world feasibility. The findings should not be generalized to physically embodied robots without further investigation, because embodiment can influence trust, social attribution, and perceived presence \cite{Wainer2007Embodiment,li2015benefit}. Future studies should isolate individual channel components,
evaluate the alternative modalities or techniques suggested
by participants, and investigate transfer to physically
embodied social robots. The qualitative findings identify possible interpretations rather than their prevalence. Because the interviews were joint and retrospective, participants may have influenced one another's accounts, and limitations of retrospective recall and subsequent disclosure may also have shaped their responses. The interviews do not provide real-time, objective evidence. Longitudinal research should examine how usability, noticing, and social interpretations develop through repeated exposure.

\section{Conclusion}

The wearable-mediated channel supported discreet interaction with a screen-based conversational agent representing a social robot within the shared game. The prescribed exchange and resulting opponent selection occurred as intended in 39 of 40 channel-mediated turns. Wearers reported low perceived
pressure/tension in both conditions, with no statistically
significant difference, and favorable immediate usability. Retrospective interviews revealed contrasting perceptions of
teamwork and unfairness, differing accounts of noticing among
non-wearers, and one wearer's reported decision to withhold
a response. These findings highlight that differential
information access can support coordination while also
creating tensions between participants' interests. Future
designs should therefore consider wearer agency, non-wearers'
awareness, and perceptions of fairness alongside wearers'
immediate usability.

\section*{Data and Code Availability}

Supplementary materials, including a discreet condition video, questionnaires, the robot GUI, the earable GUI, card tokens, game outcomes, and the in-ear wearable code, are available at: \newline \url{https://github.com/atikkhannilgar/Discreet_Interaction}.

\begin{acks}
This work was funded by Honda Research Institute Europe GmbH.
We are grateful to the participants for taking part in our study. We would like to thank Alireza Farahany, Pavankumar Chandankar, and Michael Brilka for their assistance during the study. 
\end{acks}

\bibliographystyle{ACM-Reference-Format}
\bibliography{MyBib}
\end{document}